\documentclass[aps,prl,reprint,superscriptaddress]{revtex4-1}
\usepackage[utf8]{inputenc}
\usepackage[english]{babel}
\usepackage{amsmath}
\usepackage{amsfonts}
\usepackage{amssymb}
\usepackage{graphicx}
\usepackage{url}
\usepackage[usenames]{color}
\begin{document}
\author{A.M.~Shikin}
\author{A.A.~Rybkina}
\author{I.I.~Klimovskikh}
\author{M.V.~Filianina}
\affiliation{Saint Petersburg State University, Saint Petersburg, 198504 Russia}
\author{K.A.~Kokh}
\affiliation{Saint Petersburg State University, Saint Petersburg, 198504 Russia}
\affiliation{Novosibirsk State University, Novosibirsk, 630090 Russia}
\affiliation{V.S.~Sobolev Institute of Geology and Mineralogy, Novosibirsk, 630090 Russia}
\author{O.E.~Tereshchenko}
\affiliation{Saint Petersburg State University, Saint Petersburg, 198504 Russia}
\affiliation{Novosibirsk State University, Novosibirsk, 630090 Russia}
\affiliation{A.V.~Rzhanov Institute of Semiconductor Physics, Novosibirsk, 630090 Russia}
\author{P.N.~Skirdkov}
\author{K.A.~Zvezdin}
\author{A.K.~Zvezdin}
\affiliation{Moscow Institute of Physics and Technology, Institutskiy per. 9, 141700 Dolgoprudny, Russia}
\affiliation{A.M.~Prokhorov General Physics Institute, Russian Academy of Sciences, Vavilova 38, 119991 Moscow, Russia}
\affiliation{Russian Quantum Center, Novaya St. 100, 143025 Skolkovo, Moscow Region, Russia}
\title{Spin current and magnetization induced by circularly polarized synchrotron radiation in magnetically-doped topological insulator Bi$_{1.37}$V$_{0.03}$Sb$_{0.6}$Te$_2$Se}
\begin{abstract}
We propose a hole-induced mechanism of spin-polarized current generation by circularly polarized synchrotron radiation and corresponding induced magnetization in magnetically-doped topological insulators Bi$_{1.37}$V$_{0.03}$Sb$_{0.6}$Te$_2$Se. Considered spin-polarized current is generated due to the spin-dependent depopulation of the Dirac cone topological surface states at the Fermi level and subsequent compensation of the generated holes. We have found experimentally and theoretically a relation between the generated spin-polarized current and the shift of the electrochemical potential. The out-of-plane magnetization induced by circularly polarized synchrotron radiation and its inversion with switching the direction of circular polarization were experimentally shown and theoretically confirmed.
\end{abstract}
\keywords{topological insulator, spin current}
\maketitle
\begin{figure}[b!]
\centering
\includegraphics[width=0.49\textwidth]{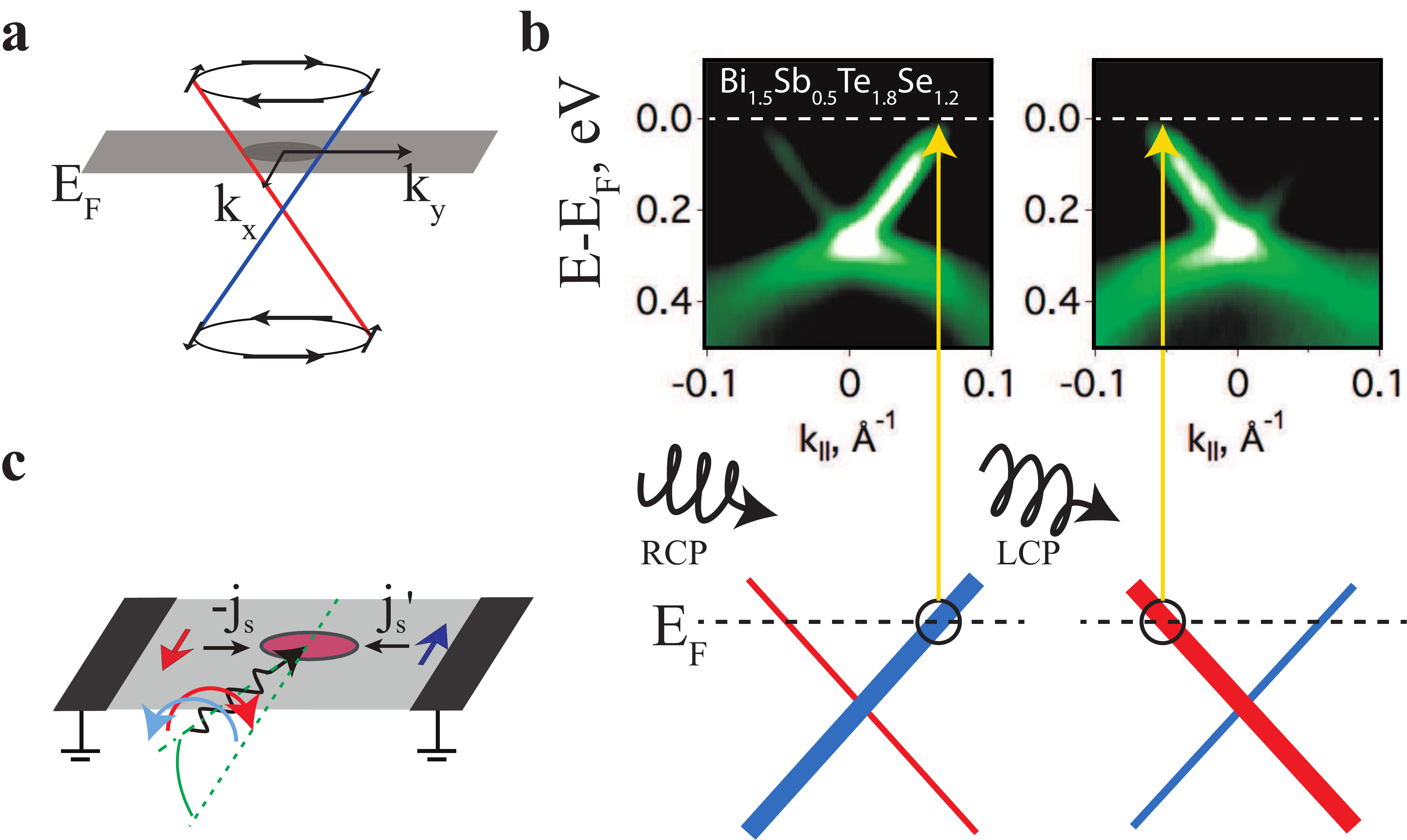}
\caption{(Color online). Schematic presentation of the Dirac cone topological surface states with helical spin structure and in-plane orientation locked perpendicular to momentum – (a) and (b) - depopulation of the Dirac cone states with opposite momentum and spin orientations at the Fermi level under photoexcitation with the right and left circular polarizations of synchrotron radiation. (c) - directions of the generated spin-polarized currents and the spin orientation under photoexcitation with opposite circular polarizations. \label{fig:1}}
\end{figure}

Topological insulators (TIs) with metallic-like topologically-protected surface states localized in the bulk energy gap are considered now among the best candidates for constructing spintronic devices with effective formation of spin currents and their controlled manipulation, see, for instance \cite{Hasan-2010, Hsieh-2009, Li-2014, McIver-2012, Xiu-2011, Xu-2012, Chen-2010}. The topological surface states (TSSs) are characterized by helical spin structure with spin locked perpendicular to momentum, see Fig.\ref{fig:1}.a. Therefore, any surface current flowing along the TSSs becomes spin polarized, and the direction of spin polarization changes to the opposite one when the direction of the current flow is reversed \cite{Li-2014}. Moreover, any imbalance in population of the TSSs with opposite spin orientation should be followed by formation of the corresponding spin-polarized current. One effective possibility of creating such imbalance is to use circularly polarized radiation. In the case of laser radiation a circularly polarized light induces selective spin-orientation-dependent electron interband transitions onto the TSSs at the Fermi level \cite{McIver-2012, Hosur-2011, Park-2012, Ando-2010}, and the corresponding spin current generation \cite{McIver-2012}. 
\par
In the current work we analyze generation of spin-polarized current induced by circularly polarized synchrotron radiation. Significantly higher photon energy in considered case in comparison with the laser \cite{McIver-2012} enables different mechanism of spin current generation. In this case a selective photoexcitation of electrons from the opposite branches of the Dirac cone states occurs in dependence on the direction of circular polarization (right- or left- one), see Fig.\ref{fig:1}.b. Due to high kinetic energy, the photoexcited electrons are emitted mainly into vacuum, and part of them are scattered with loss of the initial spin orientation. Thereat, photoexcitation is followed by generation of the long living holes at the Fermi level with momentum and spin orientations determined by the direction of circular polarization, see open circles in Fig.\ref{fig:1}.b. Since the TSSs are metallic-like and Fermi level is localized inside the energy gap the generated holes are mainly compensated by electron from the TSSs at Fermi level outside the surface region of the synchrotron beam localization. As a result, the formed compensating current flowing through the TSSs is spin-polarized as well. Because circularly polarized light excites electrons with spin oriented parallel to the light propagation the generated spin-polarized current will be oriented perpendicular to the synchrotron radiation incidence plane, see Fig.\ref{fig:1}.c. Thereat, the spin orientation in the currents is defined by the circular polarization direction (left or right). The selective generation of the holes at the Fermi level (i.e. depopulations of the TSSs) and formation of corresponding spin-polarized currents should be followed by the redistribution of the electro-chemical potential inside and outside the radiation beam spot localization, which creates the a rigid energy shift between the branches of the Dirac cone states in photoemission spectra measured at different polarizations of synchrotron radiation. The bulk and space charging effects under the action of synchrotron and laser radiation can be followed by the Fermi level shift in photoemission spectra which can be experimentally observed \cite{Zhou-2005, Hajlaoui-2014}. However, the shifts measured between the different polarizations are related to the momentum dependent depopulation of the TSS at the Fermi level and corresponding spin accumulation.
\par
A possibility of such hole-induced mechanism of spin-polarized current generation by circularly polarized synchrotron radiation applied to V-doped TI grown from the melt with a stoichiometry Bi$_{1.37}$V$_{0.03}$Sb$_{0.6}$Te$_2$Se (BSTS:V) is investigated in this work. Similar kind TIs with fractional stoichiometry (without V-doping) are characterized by effective spin transport properties due to reduced influence of bulk carriers and significant contribution of the TSS into the surface conductivity \cite{Ren-2011, Tang-2013, Wang-2015}, which assumes an enhanced efficiency of surface spin-polarised current generation in such compounds. In our work we experimentally confirm a possibility of generation of spin-polarized current by circularly polarized synchrotron radiation through the energy shift between the Dirac cone branches in photoemission spectra measured at opposite circular polarizations and through the corresponding induced magnetization reversal of magnetic dopants. For testing of the induced magnetization we have measured the modification of the out-of-plane spin structure of BSTS:V at the Dirac point under photoexcitation with different polarizations. These measurements prove that circularly polarized synchrotron radiation can actually induce a reversal magnetization in this compound with switching the out-of-plane spin orientation in the region of the Dirac point. The effects of magnetization reversal induced by the spin-polarized current generated due to the electrical field applied along the surface are well known for ferromagnetic (FM) layer deposited on Bi$_2$Te$_3$ \cite{Li-2014, Mellnik-2014} and for other systems with enhanced spin-orbit coupling \cite{Miron-2010, Miron-2011, Shikin-2014, Manchon-2009, Hals-2013, Checkelsky-2012, Chernyshov-2009, Fan-2014, Liu-2012}. The magnetization induced in TIs by synchrotron radiation was not studied yet.
\par
\begin{figure}[h!]
\centering
\includegraphics[width=0.49\textwidth]{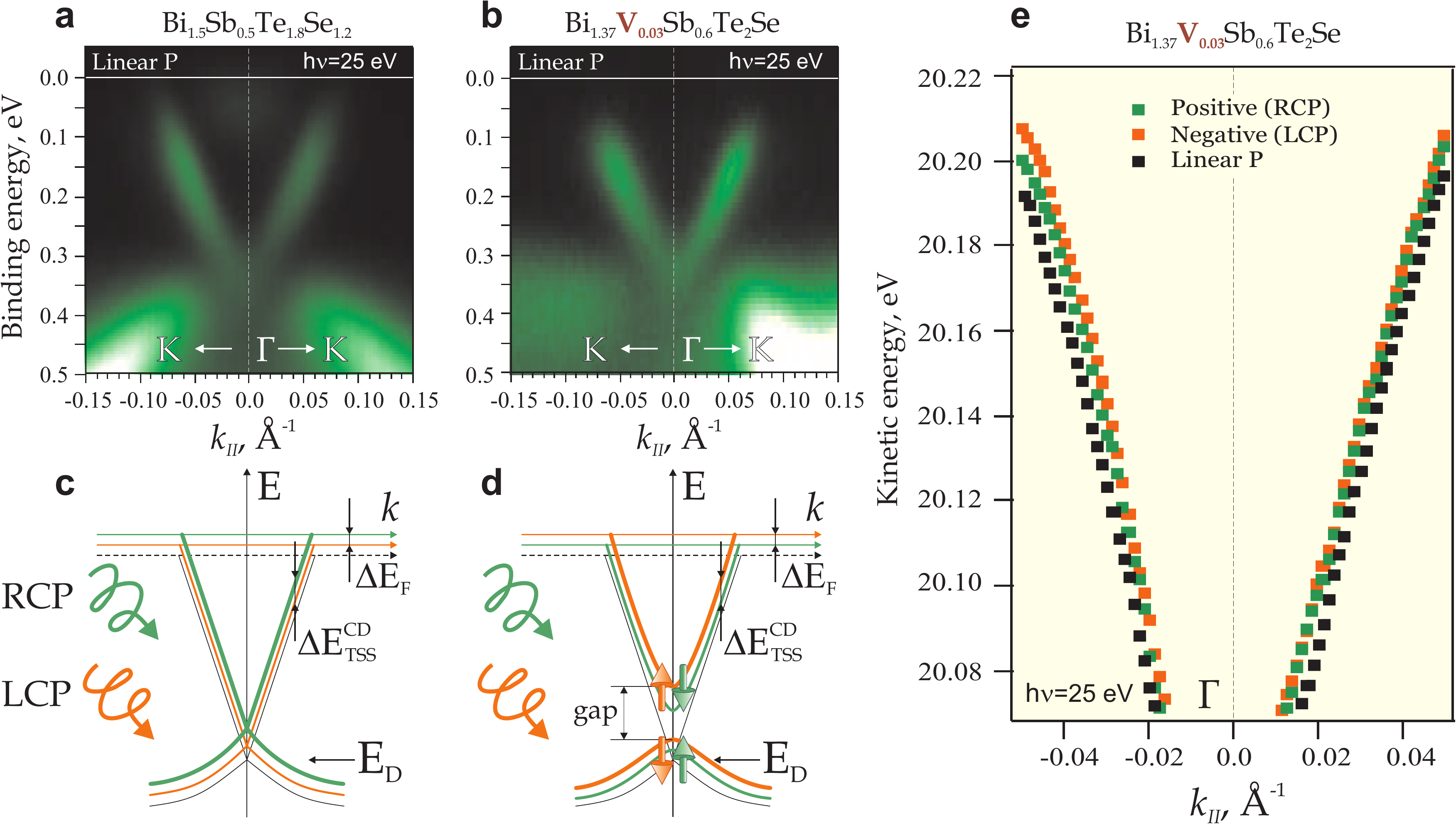}
\caption{(Color online). (a,b) – the dispersions of the Dirac cone TSSs for Bi$_{1.5}$Sb$_{0.5}$Te$_{1.8}$Se$_{1.2}$ and Bi$_{1.37}$V$_{0.03}$Sb$_{0.6}$Te$_2$Se, respectively, measured in the $\Gamma K$ direction of the BZ at $hv=25~eV$ using linear polarization of synchrotron radiation.  (c,d) - schematic presentation of the energy shift between the Dirac cone branches for pure and magnetically-doped TIs, respectively, and opening the gap at the Dirac point appeared under photoexcitation at opposite circular polarization. The reversible out-of-plane spin polarization at the Dirac point developed at opposite circular polarizations is shown by orange and green colors. (e) - the energy shift between the TSS branches estimated by fitting Lorenzian peak shapes to the spectra of the upper Dirac cone measured at different polarization of synchrotron radiation (opposite circular and linear ones). \label{fig:2}}
\end{figure}

For indication of the features of electronic structure Figs.\ref{fig:2}.a,b show the dispersions of the Dirac cone TSSs for BSTS:V in comparison with that for Bi$_{1.5}$Sb$_{0.5}$Te$_{1.8}$Se$_{1.2}$ (BSTS), which were measured in the $\Gamma K$ direction of the Brillouin zone (BZ) at $hv=25~eV$ using linear polarization of synchrotron radiation. For linear polarization an intensity of the TSSs at opposite $k_\parallel$ should be equivalent. The photoemission spectra measured at i3 beamline (MAXlab) with the synchrotron radiation incidence angle of $73^\textbf{o}$ relative to the surface normal using a Scienta R4000 analyzer. For reducing a charging effect the surface of the sample was directly grounded at the edges. The investigated monocrystals were synthesized using a modified vertical Bridgman method \cite{Kokh-2014}. The presented electronic structures for both compounds are practically similar that assumes analogous efficiency of the spin current formation. The Dirac points are located at $0.32-0.35~eV$ Binding Energy (BE). The Fermi level is arranged inside the bulk energy gap without any visible contributions of the conduction band states at the Fermi level. It allows assuming a high efficiency of development of compensating spin-polarized current flowing along the TSSs without shunting influence of the conduction band states. Fig.\ref{fig:2}.c shows a schematic presentation of the energy shift between the TSS branches of TIs appeared under photoexcitation at different polarizations of synchrotron radiation assuming corresponding redistribution of electro-chemical potential under generation of the holes at the Fermi level.
\par
Fig.\ref{fig:2}.e plots the experimental energy shift between the Dirac cone TSS branches measured for BSTS:V at different polarization of synchrotron radiation (opposite circular and linear ones) at $hv=25~eV$. The shifts between the energy positions of TSSs (at the same $k_\parallel$) measured at positive (negative) circular and linear polarizations are related to the redistribution of electro-chemical potential and corresponding generation of the induced spin-polarized currents compensating the holes created by circularly polarized synchrotron radiation. The energy positions of the TSSs presented in Fig.\ref{fig:2}.e were estimated by fitting Lorenzian peak shapes to the spectra of the upper Dirac cone in order to track their precise energy position in photoemission spectra at different polarizations. We can see evident energy shift between the measurement at opposite circular and linear polarizations.
\par
Let us estimate the spin-polarized currents generated by circularly polarized synchrotron radiation. As we noted before, photoexcitation by circularly polarized radiation creates selective imbalance in depopulation of the TSSs with opposite momenta. It results in corresponding steady-state imbalance in concentration of holes at the Fermi level characterized certain orientation of spin directed along the photon wave vector. More quantitatively this imbalance is determined by a probability of generation of the hole under photoexcitation with circular polarization in unity of time – $P(t)$ and a recombination time of the created holes – $\tau$.
\par
One can estimate the generated surface spin-polarized current in the simplest approximation as $j_s\approx eP(t)\tau V_h$, where $V_h$ is the velocity of the holes at the Fermi level related to the Dirac point position. Using the simplest form of Hamiltonian for TSSs $\hat{H}=\hbar V_D[\vec{k}\times\vec{\sigma}]\vec{z}_0$ the holes velocity can be represented as $\vec{V}_h=V_D[\vec{s}\times\vec{z}_0]\frac{2}{\hbar}$, where $\vec{s}$ is the spin of the hole and $\vec{z}_0$ is a unit vector perpendicular to the surface. Using it surface spin-polarized current takes form:
\begin{equation}
\label{eq:js}
j_s\approx\pm eP(t)\tau V_D,
\end{equation}
where $\pm$ is determined by a chirality of light. The value $P(t)$ can be calculated as sum of all photoemission matrix elements between the initial and final excited states depending on photon energy and on the angle of the synchrotron radiation incidence. At the same time the value $P(t)\tau$ is the steady-state concentration of the holes. Taking into account the fact, that Fermi level is slightly shifted ($\Delta E_F\ll E_F$) inside the beam, one can estimate $P(t)\tau$ as:
\begin{equation}
\label{eq:N}
P(t)\tau=\sum \limits_{\vec{k}}\hat{n}_k=\frac{1}{(2\pi)^2}\int_{E'_F}^{E_F}d\vec{k}=\frac{E_F}{2\pi\hbar^2V_D^2}\Delta E_F,
\end{equation}
where $\Delta E_F=E_F-E'_F$ is the measured energy shift of the Fermi edge due to the TSS depopulation under photoexcitation by circularly polarized light (see the shift between the TSS branches in Fig.\ref{fig:2}.e). If we use experimental values of $E_F\sim 0.3~eV$, $\Delta E_F\sim 2-5~meV$ (in average) and $V_D\sim 6.9\times 10^6~cm/s$, the simplest estimations give approximate magnitude of the generated spin-polarized current on the level of $j_s\sim 3-10~A/cm$.
\par
For more direct experimental confirmation of the spin-polarized current generation we have analysed the magnetization in BSTS:V induced by circularly polarized synchrotron radiation. According to Refs.\cite{Miron-2010, Miron-2011} spin-polarized current produced by a voltage applied along the interface Pt/Co/AlO$_\mathbf{x}$ induces the reversible out-of-plane magnetization in the magnetic Co dot. In our case the spin current is generated by circularly polarized synchrotron radiation. The effects of the modification of electronic and spin structure of magnetically doped TIs were investigated in Refs.\cite{Xu-2012, Chen-2010, Checkelsky-2012, Wray-2011}. It was established that the induced magnetization is followed by opening the energy gap at the Dirac point due to the time reversal breaking. It results in an opposite out-of-plane spin polarization at the opposite borders of the energy gap \cite{Xu-2012, Wray-2011} assuming a switching the out-of-plane spin structure under reverse of the developed magnetization \cite{Chernyshov-2009, Fan-2014}. Taking it into account we assumed that generation of spin-polarized current by circularly polarized synchrotron radiation in magnetically doped TI Bi$_{1.37}$V$_{0.03}$Sb$_{0.6}$Te$_2$Se can lead to the induced out-of-pane magnetization of the diluted V-atoms. It should manifest itself in formation of the out-of-plane spin polarization of the TSSs at the Dirac point with the spin splitting at the borders of the energy gap.
\par
Fig.\ref{fig:2}.d shows a schematic presentation of the modification of electronic structure of the TSSs expected for magnetically doped TI under photoexcitation by circularly polarized radiation including the energy shift between the TSSs branches discussed above and opening the gap at the Dirac point due to the induced out-of-plane magnetization. The thin black lines present the simplest structure without induced magnetization. Switching between opposite circular polarizations (orange and green color) leads to reversing out-of-plane spin polarization at the $\Gamma$ point. It is shown by corresponding arrows at the opposite borders of the gap at the Dirac point. For nonmagnetic TIs the spin structure should be degenerated at the Dirac point, see Fig.\ref{fig:2}.c.
\par
\begin{figure}[h!]
\centering
\includegraphics[width=0.49\textwidth]{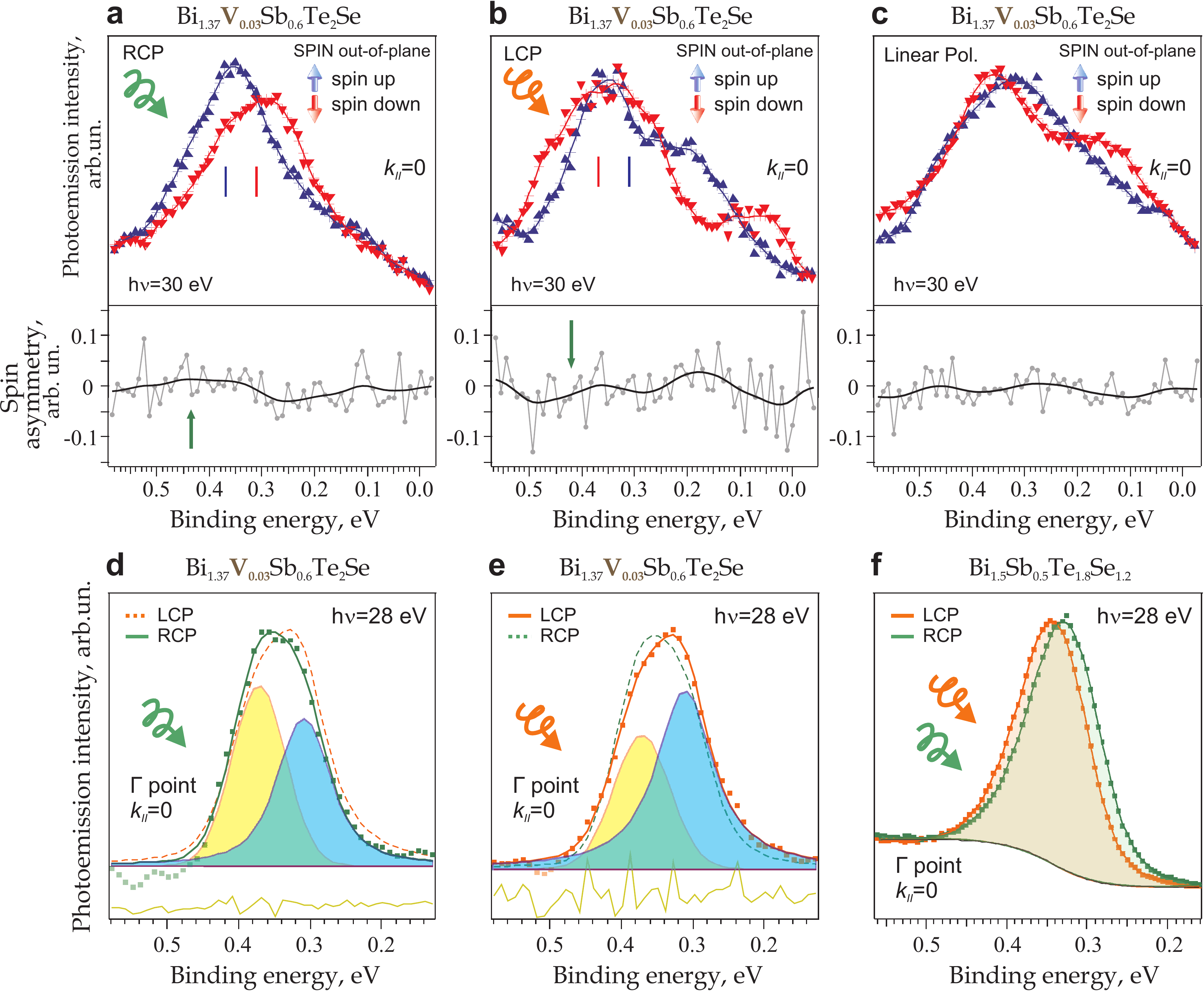}
\caption{(Color online). (a, b,c) - spin-resolved photoemission spectra measured at room temperature for V-doped TI (Bi$_{1.37}$V$_{0.03}$Sb$_{0.6}$Te$_2$Se) at the Dirac point (at $k_\parallel=0$) at $hv=30~eV$ in the direction of the out-of-plane spin polarization with using opposite circular and linear p-polarizations, respectively, showing the reversal out-of-plane magnetization induced by circularly polarized synchrotron radiation and its absence at linear polarization. Corresponding experimental asymmetries are shown below the spin-resolved spectra, too, for comparison. (d,e) – photoemission spin-integrated EDC profiles measured at $k_\parallel=0$ in the region of the Dirac point for opposite circular polarizations with the decomposition on two spectral components. (e) – similar photoemission spin-integrated EDC profiles measured at $k_\parallel=0$ in the region of the Dirac point for TI without magnetic dopants (Bi$_{1.5}$Sb$_{0.5}$Te$_{1.8}$Se$_{1.2}$). \label{fig:3}}
\end{figure}
Figs.\ref{fig:3}.a,b plot the spin-resolved photoemission spectra for BSTS:V at the Dirac point ($k_\parallel=0$) measured for the out-of-plane spin orientation under photoexcitation at opposite circular polarizations. The spectra were measured at $hv=30~eV$ that allowed to increase the intensity of the features in the region of the Dirac point. The presented spectra show a pronounced out-of-plane spin polarization including spin splitting of the states at borders of assuming gap at the Dirac point. Corresponding experimental asymmetry between the measurements at opposite circular polarizations are presented below the spin-resolved spectra. Under switching between right and left circular polarization of synchrotron radiation the spin splitting is reversed. It is interesting that under photoexcitation with linear p-polarization the spin-resolved spectra do not show noticeable out-of-plane spin splitting and polarization of the states in the region of the Dirac point, see Fig.\ref{fig:3}.c. The spectra were measured at room temperature. In relation to it, we have to note that photoexcitation by circularly polarized radiation and formation of spin-dependent imbalance in depopulation of the TSSs at the Fermi level should take place independently on temperature, of course, with different relaxation time. Therefore, the induced magnetization should lead to in-time formation of the out-of-plane spin polarization at the Dirac point including at room temperature.
\par
To confirm independently the modification of the structure induced by circularly polarized synchrotron radiation we have measured the spin-integrated energy distribution curves (EDCs) at $k_\parallel=0$ at opposite circular polarizations for the states in the region of the Dirac point. These EDCs are presented in Figs.\ref{fig:3}.d,e by green and orange colors. The fitting procedure shows a decomposition of the spectra on two components with the energy separation approximately similar to the spin splitting presented in Figs.\ref{fig:3}.a,b. The intensities of the components are reversed for opposite circular polarizations. It means that the modification of the spin structure at the Dirac point under photoexcitation with opposite circular polarization of synchrotron radiation manifest itself also in the spin-integrated spectra. For comparison Fig.\ref{fig:3}.f plots analogous EDCs measured for Bi$_{1.5}$Sb$_{0.5}$Te$_{1.8}$Se$_{1.2}$. The EDC profiles measured for this TI without magnetic dopants cannot already be decomposed on two components.
\par
The observed experimentally reversal out-of-plane spin polarization at the Dirac point is directly related to the reversal out-of-plane magnetization of V-atoms induced by circularly polarized synchrotron radiation. We connect the induced magnetization to the spin-dependent depopulation of the TSSs developed at opposite circular polarizations and corresponding uncompensated spin accumulation related to the spin-polarized current formation. The theoretical estimations for the effects described above can be follows. The spin-polarized current generated by circularly polarized synchrotron radiation (see Eq.\ref{eq:js}) induces a spin accumulation in the direction perpendicular to the surface. This accumulation can be estimated as $\delta \vec{S}=\frac{\hbar}{2}P(t)\tau\eta\vec{n}$, where $\eta=\pm 1$ is determined by a chirality of light and $\vec{n}$ is a unit vector along the wave vector of the radiation. Using value of the holes concentration from Eq.\ref{eq:N} the accumulation takes form:
\begin{equation}
\label{eq:S}
\delta S_z=\frac{E_F\eta n_z}{4\pi\hbar V_D^2}\Delta E_F,
\end{equation}
where $n_z$ is the projection of $\vec{n}$ on axis perpendicular to the surface. The induced spin accumulation breaks a time reversal symmetry induces energy gap $\Delta$, hence induces effective out-of-plane magnetic field $h_{eff}=\Delta/2\mu_B$ which creates the out-of-plane magnetization of V atoms $\delta S_z=\frac{\hbar}{2\mu_B}\chi_P h_{eff}$, where $\chi_P=\frac{4\mu_B^2}{E_F}n$ is Pauli magnetic 2D-susceptibility of TSSs and $n$ is the concentration of electrons at TSSs with corresponding position of the Fermi level. Using Eq.\ref{eq:S} one can find the effective field in this case:
\begin{equation}
\label{eq:h}
h_{eff}=\frac{\mu_B}{\chi_P}\frac{E_F\Delta E_F}{2\pi\hbar^2 V_D^2}n_z\eta.
\end{equation}
This effective magnetic field creates out-of-plane magnetization of V-ions subsystem. The effective exchange field in this case take form of $h_{eff}$ with additional factor $\varsigma\sim 1$, which is the ratio between the exchange constants in cases of free electron – free electron and free electron – localized electron. Then the induced magnetization can be estimated:
\begin{equation}
\label{eq:m}
\delta m_z^V=\frac{\chi_V}{\chi_P}\frac{\mu_B E_F\Delta E_F}{2\pi\hbar^2 V_D^2}n_z\eta\varsigma,
\end{equation}
where $\chi_V\approx\frac{g_V^2\mu_B^2}{3kT}n_V$ is magnetic susceptibility of V ions, $n_V$ – the concentration of V-atoms. This reversible with radiation polarization out-of-pane magnetization leads to corresponding changes of the out-of-plane spin structure in the region of the Dirac point. This reverse switching of the spin structure is presented in Fig.\ref{fig:3}.a,b for opposite circular polarizations of synchrotron radiation.
\par
We have presented mechanism of spin current generation in TIs along the surface by circularly polarized synchrotron radiation due to the spin-dependent depopulation of the Dirac cone TSSs at the Fermi level and following compensation of the generated holes. It was shown that photoexcitation by circularly polarized synchrotron radiation and generation of corresponding holes are followed by the energy shift between the Dirac cone state braches for opposite direction of circular polarization related to the redistribution of the electrochemical potential between the areas inside and outside the synchrotron radiation beam localization due to the hole generation.
\par
We have shown a possibility of reversal out-of-plane magnetization induced in magnetically doped TI (BSTS:V) by circularly polarized synchrotron radiation even at room temperature. The out-of-plane spin structure induced at the Dirac point can be switched with changing the direction of circular polarization. Theoretical estimations have confirmed a possibility of the induced out-of-plane magnetization by circularly polarized synchrotron radiation and its inversion with switching the direction of circular polarization. It was found the relation of the induced out-of-plane magnetization with the spin accumulation developed due to the spin-dependent depopulation of the Dirac cone states at the Fermi level with opposite momentum and spin orientation generated by circularly polarized synchrotron radiation. The magnitude of the induced magnetization was estimated.
\par
The work was supported by grants of Saint Petersburg State University for scientific investigations (No. 11.38.271.2014 and 15.61.202.2015), RFBR Grant No. 14-02-31781 and the 50 Labs Initiative of the Moscow Institute of Physics and Technology. The authors kindly acknowledge the MAXlab staff for technical support and help with experiment and useful discussions.
\providecommand{\noopsort}[1]{}\providecommand{\singleletter}[1]{#1}%
\end{document}